\documentclass[aps,  groupedaddress,  twocolumn,  showpacs, showkeys, floatfix]{revtex4}

\usepackage{graphicx}
\usepackage{amsmath}
\begin{document}
\newcommand{\pthz}[1]{
\mbox{$ #1\;\mathrm{pT}/\sqrt{\mathrm{Hz}}$}}

\newcommand{\fthz}[1]{
\mbox{$ #1\;\mathrm{fT}/\sqrt{\mathrm{Hz}}$}}

%\preprint{}


\title{Stray Magnetic Field Compensation with a Scalar Atomic Magnetometer}

\author{J.\,Belfi }
\altaffiliation
{Currently at Dept. of Physics, University of Pisa.}

\author{G.\,Bevilacqua}
\affiliation{CNISM UdR Siena - Dept.\,of Physics and CSC, University of Siena, Via Roma 56, 53100 Siena, Italy.}

\author{V.\,Biancalana \footnote{Corresponding Author, biancalana@unisi.it}
}
\affiliation{CNISM UdR Siena - Dept.\,of Physics and CSC, University of Siena, Via Roma 56, 53100 Siena, Italy.}

\author{R.\,Cecchi}
\affiliation{CNISM UdR Siena - Dept.\,of Physics and CSC, University of Siena, Via Roma 56, 53100 Siena, Italy.}

\author{Y.\,Dancheva}
\affiliation{CNISM UdR Siena - Dept.\,of Physics and CSC, University of Siena, Via Roma 56, 53100 Siena, Italy.}

\author{L.\,Moi}
\affiliation{CNISM UdR Siena - Dept.\,of Physics and CSC, University of Siena, Via Roma 56, 53100 Siena, Italy.}

\date{\today}

\begin{abstract}
We describe a system for the compensation of time-dependent stray  magnetic fields using a dual channel scalar magnetometer based on non-linear Faraday rotation in synchronously optically pumped Cs vapour. We  detail  the active control  strategy,  with an  emphasis  on the electronic circuitry,  based on a simple phase-locked-loop integrated circuit. The performance and limits of the system developed   are tested and discussed. The system was applied to significantly  improve  the detection of  free induction decay signals from protons of remotely magnetized water precessing in an ultra-low  magnetic field.


%33.57.+c Magneto-optical and electro-optical spectra and effects

%85.70.Sq Magnetooptical devices

%07.55.Ge Magnetometers for magnetic field measurements

%07.55.Nk Magnetic shielding in instruments

%41.20.Gz Magnetostatics; magnetic shielding, magnetic induction, boundary-value problems


\end{abstract}

\maketitle

\section{Introduction}

High sensitivity magnetometry has a wide range of applications in both  fundamental and applied research. Nowadays, the best performance in terms of sensitivity is  provided by Superconducting Quantum Interference Devices (SQUIDs) and by Atomic Optical Magnetometers (AOMs) operating in the Spin-Exchange Relaxation-Free (SERF) regime. In both cases, the impressive limit of \fthz{1} has been  surpassed experimentally, when operating in volumes carefully shielded against the environmental magnetic noise. \cite{romalisnature07}

Scalar magnetometers operating in non-vanishing magnetic fields have also been reported to have a sensitivity as good as a few tens of \fthz{} when operating in shielded environments. This value  degrades to about \pthz{1} in unshielded volumes.

The precise determination and control of magnetic fields is a key factor  in atomic physics experiments \cite{DedmanRSI07, RingotPRA01, HeinzOC06}, fundamental research (such as determination of the neutron electric dipole moment  \cite {groegerEPJD06, BrysNIM05}),  ultra-low-field (ULF) NMR and imaging
\cite{savukovjmr07}, and the development of biomagnetic diagnostics \cite{yamazNCN04, resmerSST04}.

Field control is commonly attempted through both passive shielding and active compensation approaches, and many solutions are reported in the literature. Different applications require that the shielding/compensating systems have different levels of noise rejection and respond in different frequency ranges.

Passive shielding can be achieved by surrounding the measurement volume with soft-iron/large permeability material, or with thick conducting material (or superconducting layers) in which  eddy currents are induced. The superconducting approach has been studied and developed since the 1970s \cite{cabrera73, cabrera88, resmerSST04} and yields the highest field stability. 

The efficiency of passive shielding decreases at low frequency, thus shields can be profitably coupled with active compensation systems to improve the performance in the Hz and sub-Hz range. Active compensation has also been proposed and studied  for use in compensating peaked spectral components of the stray field, such as those at 50\,Hz or 60\,Hz due to power lines. Active compensation of deterministic disturbances has also been attempted  with a feed-forward approach \cite{hasegawaRSI68}. Nevertheless the most recent literature refers to feedback systems, which are suitable for attenuating both deterministic and random components.

Active compensation with a suitably extended bandwidth can also be applied in stand-alone configurations (i.e. not combined with passive shielding). This  is of interest, for instance, where easy access (no geometrical constraints) is required, and/or when large magnetically clean volumes are needed, which would make the passive shield bulky and very expensive. 

Naturally, the specifications (bandwidth and attenuation factor) and the final performance of such active set-ups depends on the feedback design and on the sensors used to generate the error signals. In the latter respect, a large variety of sensors were reported in the literature, ranging from the mechanical ones (a sort of compass!) of decades ago \cite{ScottRSI57}  to Hall sensors \cite{Peters}, magnetoresistances \cite{BottiRSI06, RingotPRA01}, fluxgates \cite{BrysNIM05, IoanSandA97}, SQUIDs \cite{skakalaMST93, PlatzekRSI99}, 
SERF-AOMs \cite{SeltzerAPL04}, NMR based magnetometers \cite{ericksonRSI80}.

A further approach frequently used to counteract magnetic noise relies on the fact that noise sources, being at a large distance from the measurement volume, produce a rather homogeneous disturbance. Thus the use of differential (gradiometric) sensors make it possible to reduce noise when measuring signals produced by nearby sources \cite{bisonAPL09}, so that  the common-mode signal can be attributed to magnetic noise and rejected. 

In this paper we present the results obtained with a dual channel scalar  Cs AOM operating in $\mu$T field, which was used to stabilize the modulus of the bias field. This set-up makes it possible to detect the low-level  time-dependent anomalies generated by small sources  located close to one of the two sensors composing the differential magnetometer. Both sensors work in the same bias magnetic field, which is also the field in which the anomaly source is immersed. The stability of the bias field plays a key role when the magnetometer is used to detect NMR free induction decay: stabilizing the field makes the nuclear precession occur at a more stable frequency, thus facilitating its analysis and long lasting averaging.

While the characteristics of the dual channel magnetometer and of its applications to ultra-low-field NMR detection \cite{noijosa09, noijmr09} and magneto-cardio signal detection \cite{noijosa07} are reported elsewhere, here we discuss the performance of a bias field stabilization system, focusing in particular on the effects of residual magnetic noise in our simplified configuration, in which a pair of scalar sensors are used to counteract the noise in the field modulus by acting on a single component.

In the case of a self-oscillating magnetometer, considered as a field-to-frequency converter, when the field compensation is controlled by a voltage, the entire setup acts as a voltage-controlled oscillator: the whole compensation system  can be regarded as belonging to a well known and extensively studied class of electronic circuits,  phase-locked loops (PLLs).  
From a practical and technical point of view, this analogy has two advantages: a complete feedback theory related to the PLL has already been developed and assessed, and the fact that a variety of integrated circuits are commercially available  simplifies the task of building high quality phase comparator circuits of several kinds, and designing loop filters. The system described in this paper uses a commercial PLL integrated circuit and  a micro-controller chip (PIC) used as a programmable divider to generate an adjustable reference frequency. We foresee that further improvements could be achieved by  using  digitally controlled PLL devices, which contain  different kinds of phase comparators and programmable loop filters \cite{PLLbook}.

\section{Experimental apparatus}
\begin{figure}
   \includegraphics[width=8cm]{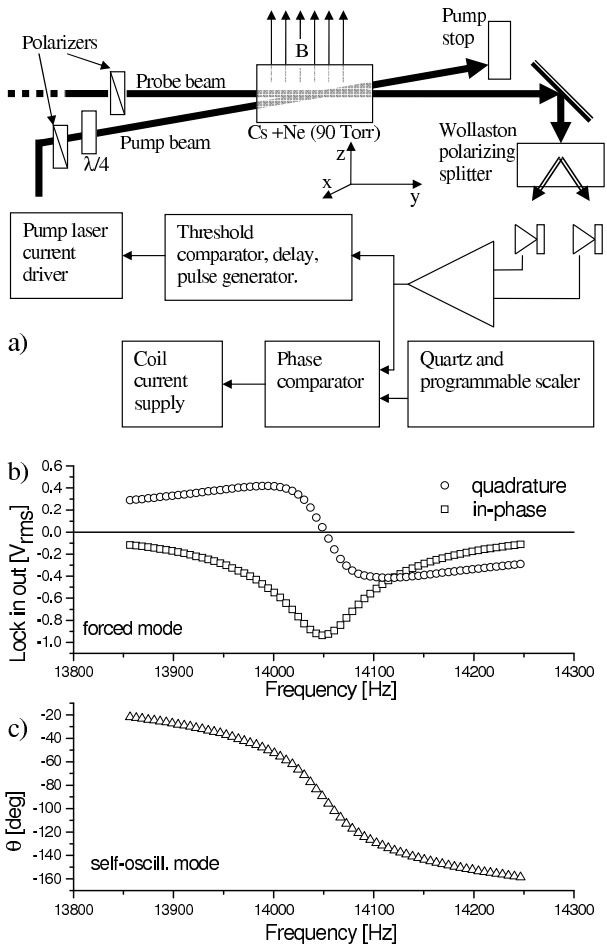}
   \caption{\label{fig:setup1arm} (a) Schematic of the self-oscillating optical atomic magnetometer (main arm). A balanced polarimeter detects the atomic precession, generating a signal that triggers the synchronous optical pumping. The same signal is phase-compared  with a stable reference signal to generate the error signal, which is fed back to stabilize the field (see Fig.\,\ref{fig:opampcirc}).  The (b) plots show the in-phase and quadrature signals obtained in scanned mode. The (c) plot shows the relationship between loop dephasing and oscillation frequency in self-oscillating mode.
         }
\end{figure}

The experimental apparatus is sketched in Figs.~\ref{fig:setup1arm} and \ref{fig:appsp}. Two sealed cells containing Cs and 90 Torr of Ne as a buffer gas are heated to about 45$ ^ \circ $  C. The two cells are displaced by 7\,cm along the $\hat x$ direction. Each cell is illuminated by two overlapping laser beams, both attenuated down to a few $\mu$W/cm$^2$. One of them (the pump beam) is circularly polarized and broadly frequency modulated  so to make it pass periodically in and out of the Doppler-broadened atomic resonance. The other beam (the probe beam) is linearly polarized and unmodulated. The bias magnetic field ($B_0 \hat z$)  is perpendicular to the beams' propagation axis ($\hat y$). Provided that the pump beam excites atoms synchronously with their Larmor precession, a precessing macroscopic polarization of atoms is induced in both the cells. The probe beam polarization undergoes a time-dependent Faraday rotation of the polarization, which is detected by means of two balanced polarimeters. The modulation frequency of the pump laser can be scanned to detect and characterize the resonance, or  be produced by a pulse generator triggered by the polarimetric signal of the main channel, as represented in Fig.~\ref{fig:setup1arm}. The latter configuration makes the system a self-oscillating magnetometer. The plots show the resonance shape detected in scanned mode and the dependence of the oscillating frequency as a function of the loop dephasing introduced as a feedback delay. The merit factor of the resonance ($Q=\nu_0/\Delta\nu_{FWHM}$) is about 130  and is consistent with the maximum slope of the lower plot ($\pi$\,rad/110\,Hz). This value was optimized by adjusting several experimental parameters such as cell temperature, laser intensities, and probe detuning. The  value achieved is consistent with other optimized results reported in the literature \cite{versh09}.  Additional details of the self-oscillating setup are available in \cite{noijosa09, noijmr09}.

\subsection{Coil dc supplies}
\begin{figure}
     \includegraphics[width=8cm]{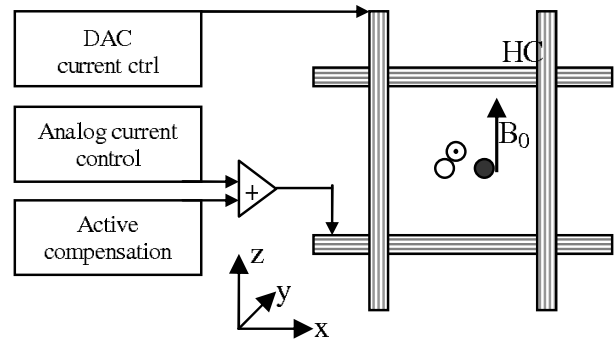}
  \caption{\label{fig:appsp} Schematic of the apparatus (not to scale). The transverse field components, $B_{x,y}$, are compensated by Helmholtz coils (HC) supplied by numerically controlled current generators. The  $B_z$ component (bias field) is analogically adjusted and actively stabilized. The symbols $\bullet$ and $\circ$  represent the  cells of the main arm and secondary arm, respectively, while  $\odot$ represents the sample. For the sake of clarity, one of the transverse HCs, and the gradient compensation quadrupoles are not shown. }
\end{figure}

Fig.\,\ref{fig:appsp} represents (not to scale) the relative positions of the atomic sensors and compensation coils. A set of three mutually orthogonal, large (180\,cm sides) Helmholtz pairs are used to control the magnetic field. Two pairs are used to cancel out the local magnetic field in the horizontal plane ($x$ and $y$ pairs), while the third pair is used to partially compensate the $z$ component, thus setting the value of the bias field to about 4 $\mu$T. Three separate current generators are used to supply the compensation coils. The current supply for the $z$ pair is analogically  controlled by high quality potentiometers, while the currents for the $x$ and $y$ pairs are digitally (and automatically) controlled.
The procedure used to zero the transverse components of the field is based on the minimization of the self-oscillating frequency: the PLL circuit is deactivated and a simplex procedure adjusts the currents in the $x$ and $y$ coil pairs while evaluating and minimizing the atomic Larmor frequency.

\subsection{Active phase-locked loop compensation}
The analogue signal generated by one balanced polarimeter is preamplified, filtered and compared to a threshold value, in order to produce a synchronous square wave which triggers (possibly after frequency scaling) the self-oscillating magnetometer. The same signal is phase-compared with a stable reference signal in order to produce the error signal for the PLL. The phase comparison is performed by  phase-comparator-1 (PhC-1, exclusive or) or by  phase-comparator-2 (PhC-2, edge-triggered three-state circuit) of a CD4046 integrated circuit.

\begin{figure}
  \includegraphics[width=8cm]{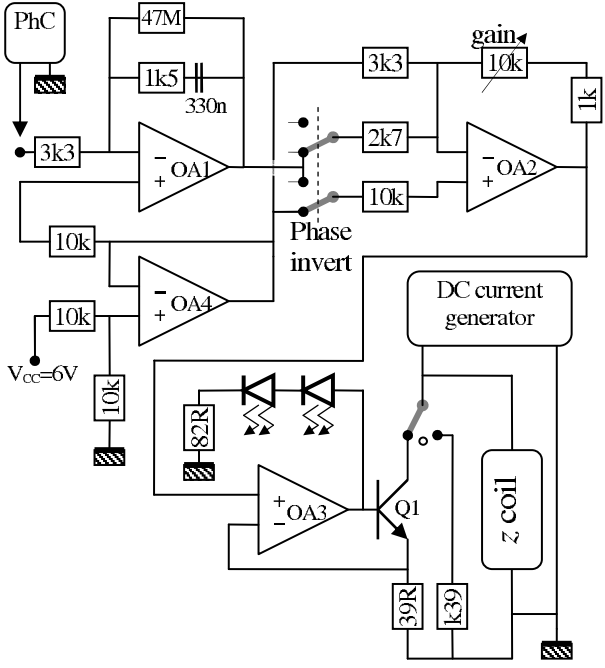}
  \caption{\label{fig:opampcirc} Schematic of the loop filter and gain adjustment circuit. The phase comparator (PhC) signal is filtered by OA1, amplified (and optionally inverted) by OA2, and converted to current by OA3 and Q1. OA4 supplies a local reference biased 3V above the ground. Two green LEDs are used as a monitor for the Q1 work-point. The Q1 collector sinks part of the Helmholtz coil current, when the closed-loop operation is selected.}
\end{figure}

A chain made of a one-pole/one-zero active filter followed by an adjustable-gain linear amplifier and by a voltage-to-current converter, closes the loop, sinking a small part of the current supplying the $z$ Helmholtz pair, see Fig.~\ref{fig:opampcirc}. 

\section{Performance analysis}

\begin{figure}
     \includegraphics[width=8cm]{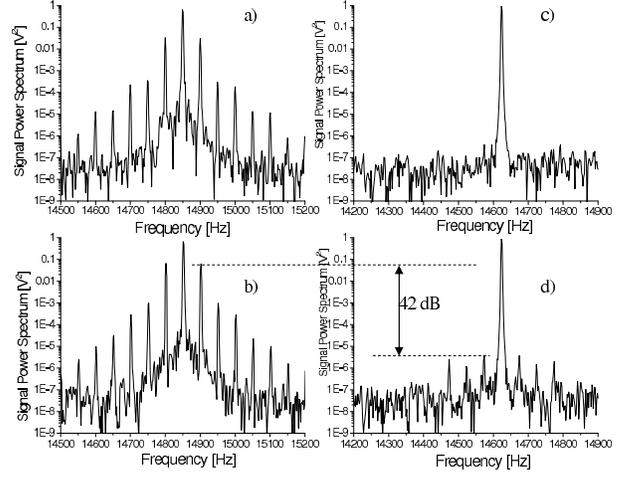}
  \caption{\label{powerspectrum} Power spectrum of the polarimetric signal. In (a) and (b) no compensation is used and 50\,Hz noise sidebands are clearly visible, with an amplitude of about $-10 \div -13$\,dB$_{\rm c}$. The plots (c) and (d)  are obtained with active compensation.  Plots (a) and (c) refer to  the main arm, while (b) and (d) refer to the secondary one. Plots (a) and (c) show that the 50\,Hz sidebands disappear, meaning that an extinction of 60\,dB is attained in the main arm. In the secondary arm (plots (b) and (d)) the 50\,Hz sidebands are attenuated by  42\,dB. Thanks to the rejection of low frequency noise, in  traces (c) and (d) the carrier peak emerges from a much weaker pedestal with respect to (a) and (b).}
\end{figure}

The self-oscillating magnetometer converts the modulus of the magnetic field to frequency, so that any modulation in the modulus appears as a pair of sidebands in the power spectral density of the magnetometric signal. The left plots in Fig.\,\ref{powerspectrum} show typical spectra recorded without any shielding or active compensation. Strong sidebands appear at the power line frequency and its multiples, as a result of significant magnetic noise produced by transformers and other electrical devices. The amplitude of such sidebands varies in time, and in the case of the  spectra shown  amounts to -10 and -13\,dB${\rm _c}$ for the main and secondary arm, respectively. The right plots in Fig.\,\ref{powerspectrum} were recorded in the same conditions, apart from having activated the compensation system:  the 50\,Hz sidebands are reduced to the broadband noise level (i.e. by about 60\,dB) in the case of the main arm and by 42\,dB for the secondary arm. In addition, for both the arms the carrier peaks emerge from much weaker pedestals, indicating a significant rejection of low frequency noise. A quantitative analysis of such noise reduction is made in the following.

\subsection{Short-term} 

The analysis of the short term performance consists in the characterization of the residual phase noise in the polarimetric  signals. The relatively low-frequency operation of the magnetometer makes  the direct digitization of these signals possible, as well as the implementation of a full numerical procedure for  phase noise evaluation.

The data analysis reproduces, using numerical tools, the typical setup  of double balanced mixers (DBM), which are commonly used in the phase noise characterization of RF and microwave devices (see \cite{rubiola, rubiola_book} for an overview and  more in depth discussion). Pair of samples are digitized at 50kS/s (or less) with an amplitude resolution of 16 bit. Each sample is a trace containing $2^{13}=8192$ data points. In these traces two out of the three signals available are recorded: the main arm polarimetric signal, the secondary arm polarimetric signal, and the PLL reference. Consequently, noise characterization can be achieved for the relative phase of the two arms or for the absolute phase of each individual arm.

Once the digitized traces are transferred to the computer, the main single tones of the two traces are evaluated by means of a Lanczos-Grandke procedure \cite{lanczos56, grandke83}, in order to extract their frequency and relative phase and to normalize their amplitude. Both traces are then interpolated by means of FFT/zero-padding/FFT$^{-1}$ technique, up to an apparent sampling rate of about 1MS/s or more.  The two digitized signals are then relative phase shifted up to a dephasing of $\pi/4$. The precision of the phase shifter is limited by the sampling rate $1/dt^\prime$ of the interpolated trace ($\delta\phi=2\pi \nu_0 dt^\prime$). This amount sets an upper limit to the accuracy of the phase measurement described below. The resulting quadrature traces are then multiplied element-by-element. The procedure described so far implements the phase-shifter and the mixing circuit of the DBM. The low pass filter which follows the mixer in DBM setups is replaced here by a single tone identification and subtraction routine, tuned to the second harmonic of the  previously detected frequency. The residual signal can thus be modeled as 

\begin{equation}
\begin{split}
    g(t)=& \sin [2\pi \nu_0 t + \phi_1(t)+\delta\phi   ] \cos  [2\pi \nu_0 t + \phi_2(t) ] \\=  
    & \frac{1}{2} (\sin[2 \times 2\pi \nu_0 t + \phi_1(t)+\phi_2(t)+\delta\phi]  \\  & + \sin[ \delta\phi+\phi_1(t)-\phi_2(t)]  ) 
\end{split}
\end{equation}
which, after having filtered out (actually subtracted) the component at $2\nu_0$, can be approximated as

\begin{equation}
        g(t)\approx \frac{ \cos(\delta \phi)\left(\phi_1(t)-\phi_2(t)\right )}{2} 
    \approx
    \frac{ \phi_1(t)-\phi_2(t)}{2} = \frac{ \Delta\phi(t)}{2} 
\end{equation}
where the two approximations consist in the linearization of  $\sin ( \Delta \phi)$ in the hypothesis $\Delta \phi=\phi_1-\phi_2 \ll 1$, and in the assumption that $\delta\phi=0$, respectively.  The phase difference $\Delta \phi(t)$ is finally analyzed by repeated FFT evaluations  of its power spectral density, which are then averaged to highlight the main features of the spectrum. The above mentioned quadrature error $\delta\phi$ implies a systematic error $(1- \cos \delta \phi)^2 \approx \delta\phi^4/4$ on the spectrum scale, which is negligible provided that the interpolation produces a good level of oversampling, e.g. we used  $dt'\approx 1/(100 \nu_0) \Rightarrow \delta\phi\approx 60$\,mrad.

\begin{figure}
  \includegraphics[width=8cm]{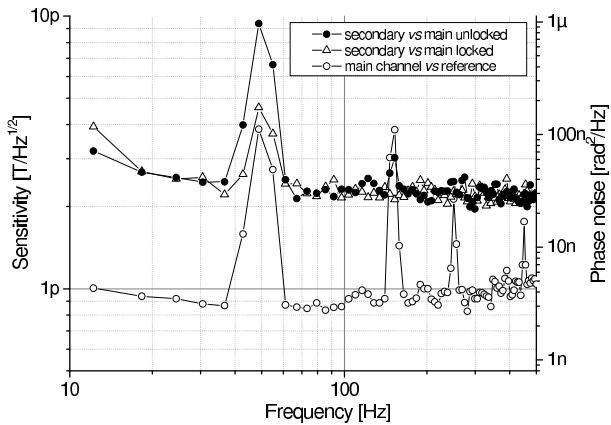}
  \caption{\label{phasenoise} Residual absolute phase noise in the main arm ($\circ$), and relative phase noise between the two arms in locked ($\triangle$) and unlocked ($\bullet$) field conditions (see text for details). The right and left vertical axes refer to phase noise and magnetic sensitivity, respectively. The conversion between them follows eq.\,\ref{phi2field}. 
  }
\end{figure}

The phase noise spectra recorded by comparing different couples of traces are plotted in  Fig.\,\ref{phasenoise}. Absolute  phase noise  in the main arm ($\circ$)  appears together with the relative phase noise of the two arms, recorded in locked-field ($\triangle$) and unlocked-field ($\bullet$) conditions (right axis units). The relative phase plots are used to infer  the differential magnetometric sensitivity, which is evaluated by converting the power spectral density of the phase noise to magnetic field units (left axis units):

\begin{equation} 
\label{phi2field}
\sqrt{S_B}=
\frac{\partial B}{\partial \varphi}\sqrt{S_\varphi}=
\frac{\nu_0}{\gamma Q}\sqrt{S_\varphi}=
\frac{\Delta \nu_{FWHM}}{\gamma }\sqrt{S_\varphi}
\end{equation}
where $\gamma  \approx  21.97 \times 10^{9}$ (rad/s)/T is the gyromagnetic factor of Cs.

The results shown in Fig.\,\ref{phasenoise} demonstrate an ultimate limit of the residual noise of less than \pthz{1} in the main arm, and of \pthz{2.5} in the secondary one. Concerning the noise at 50\,Hz,  residuals of \pthz{4} and \pthz{10} appear in the two arms, respectively. This evaluation is consistent with the rougher estimate derived from Fig.\,\ref{powerspectrum}. It is important to point out that the 50\,Hz noise varies  in time, thus  preventing accurate comparisons.
The peak at 150\,Hz is not a third-order sideband, but rather a consequence of the 50\,Hz noise anharmonicity, and is only partially attenuated in both arms as it is located close to the cut-off frequency of the loop filter.
A spurious  peak at 100\,Hz, is also occasionally observable in the phase noise of both arms. This cannot be explained in terms of a second order sideband (as it is definitely larger than estimate $J_2^2(M)/J_1^2(M)\approx M^2/16$ at the modulation index $M$ inferred from the first sideband amplitude). This is, for example, the case of the Fig.\,\ref{powerspectrum} (d). 

The  phase noise spectra can be compared to the noise level expected from the photodetectors and photocurrent amplifiers. The photocurrent shot noise, transimpedance Johnson noise, and preamplifier input current noise all contribute  with similar amounts, resulting in about $135\mu {\rm V / \sqrt{Hz}}$ at the output of each arm. This estimate, with  an   in-phase signal $V_x\approx$1\,V, leads to a white  phase noise as large as $\delta \varphi^2 \approx \delta V_y^2/V_x^2 \approx 20{\rm \, n rad^2 / {Hz}}$, consistently with the values that appear in Fig.\,\ref{phasenoise}. 

The stabilization system reduces such phase fluctuations by a factor of 4 in the main arm. This is achieved by the desired and appropriate reaction to the magnetic noise and to an unwanted reaction  to other noise sources. The latter is at the expense of an increase in the magnetic noise in the secondary arm. It is worth noting that no additional noise is introduced, as demonstrated by the substantial agreement between the plots ($\bullet$) and ($\triangle$) in Fig.\,\ref{phasenoise}. In short, the common mode noise represented by magnetic fluctuations is counteracted by the system, while phase fluctuations originating from other noise sources are compensated by magnetic field adjustments in the main arm. In this way, part of the noise detected in the main arm is transferred to the secondary arm through an additional magnetic noise.

The data shown in Fig.\,\ref{phasenoise} and in Fig.\,\ref{powerspectrum} were obtained using as a phase-comparator the PhC-1 (exclusive or). We observed that, in spite of a larger residual noise at $2 \nu_0$ in the error signal (partially transferred to the coil current), PhC-1 gave a lower background in the phase noise than PhC-2. This feature is consistent with the well known greater sensitivity to noise of PhC-2. A cleaner polarimetric signal would probably result in favourable conditions for the use of a PhC-2 instead.

\subsection{Long term}
Even in ideal conditions, a test of the long term stability of the  compensation system described, cannot be performed by analyzing the drift or the Allan variance of the self-oscillating frequency: such a test would instead characterize the relative stability of the reference oscillator and the data acquisition clock.

Evaluation of the long term stability, i.e. the amount of the noise contribution in the sub-Hz $\div$ sub-mHz range, requires  knowledge of the extent to which  a certain value of the (locked) oscillation frequency corresponds to a given value of the magnetic field. In fact, while the  self-oscillation frequency is ideally set by the Larmor frequency of atomic spins and thus by the modulus of the magnetic field, several subtle effects may induce slight deviations.  These effects have been  studied intensively as they are commonly considered as sources of systematic errors in precise magnetometry. Should they  vary in time due to drifts in the atom-light interaction conditions, an apparent field drift would appear and  would be inopportunely compensated by the active stabilization system. In conclusion, the long term stability would be set by the drifts of the systematic errors. 

We point out several potential causes of drifts of systematic errors, among which the most relevant are the cell temperature and the laser junction currents and temperatures, with their well known effects on  laser tuning.  In particular, drifts in pump laser detuning result in slight phase deviation of the synchronous pumping, thus producing drifts in the self-oscillating frequency, in accordance with the behaviour shown in Fig.\,\ref{fig:setup1arm}, plot (c). 

The relationship between loop dephasing and oscillation frequency is a delicate feature of self-oscillating magnetometers \cite{budkerrobust}. The loop dephasing is set by a time delay in the loop, whose conversion to dephasing depends on the operating frequency. For this reason, stabilizing the frequency makes the magnetometer more reliable and robust.

It is worth noting that the long term drifts limit the performance of the magnetometer to a greater or lesser extent depending on the specific application. We considered applications in magnetocardiography \cite{noijosa09} and in ULF-NMR remote detection \cite{noijmr09}. In the first case only drifts occurring during a cardiac epoch are relevant, which are definitely negligible compared to the signal dynamics.
In the case of NMR detection, the low signal-to-noise ratio makes long lasting averages necessary, so  appropriate measurement requires that the field drifts occurring during the whole average process causes the nuclear precession frequency to drift negligibly compared to the NMR resonance width. We considered averaging intervals ranging from several minutes to several hours, and the  nuclear resonance detected (at $\nu_{n}\approx$180\,Hz) had a width of $\Delta \nu_n = 1 / (2 \pi T_2) \approx$100\,mHz.
The  stability required can thus be written as $\Delta B/B=\Delta \nu_n / \nu_n \ll 5 \times10^{-4}$, to prevent an apparent reduction of $T_2$ due to inappropriate averaging.

The latter consideration suggests that  nuclear resonance can be used as an independent and  accurate magnetometer to study the long term performance of the active compensation system in terms of Allan variance of the  nuclear precession frequency. This procedure requires  the  $\nu_n$ evaluated to be determined with a very small margin of uncertainty $\delta\nu_n$. In  Sec.\,\ref{example} we report a practical example, where the best fit procedure gives a relative frequency uncertainty of $\delta\nu_n/\nu_n\approx 1.6\times10^{-6}$, meaning that the relative field drifts in this  amount, occurring in the averaging time (2800\,sec), are appreciable. The relatively poor signal-to-noise ratio of our setup makes this  procedure difficult. Nevertheless it can be proposed as a characterization method for evaluating  performance over the very long term, especially in conjunction with more sensitive magnetometers.

\subsection{Residual noise due to field inhomogeneities}
\label{residui}
As discussed above, our system allows for complete control of all the dc components of the magnetic field, while tracking only its modulus over time, and counteracting its variations through adjustments of the strongest ($z$) component. 
It is worth considering the limitations deriving from such a configuration, with the specific aim of evaluating the critical features that occur when the system is used in a dual channel setup. 

Let us assume that the system perfectly compensates the field fluctuations in the main arm, and let us derive the residual noise in the secondary arm. We will assume that the noise $\vec B_n$ and the compensation field $B_c \hat z$ are homogeneous, while the bias field may  differ due to static gradient: the field is $\vec B_0$ in the main cell, and $\vec B_1$ in the secondary one, with $\vec B_1 \approx  \vec B_0 + {\bf J} \Delta \vec x$,  ${\bf J}$ being the Jacobian of the field and $\Delta \vec x$  the  position of the secondary cell with respect to the main one. Thus, perfect compensation in the main cell reads
\begin{equation}
\label{B0}
\left (\vec B_0+ \vec B_n + B_c\hat z \right )^2= \vec B_0^2 
\end{equation}
while, concerning the secondary cell, 
\begin{equation}
\label{B1}
\left (\vec B_0+ \vec B_n + B_c\hat z  + {\bf J} \Delta \vec x \right)^2= \vec B_1^2 + \Delta(B_1^2),
\end{equation}
$\Delta(B_1^2)$ representing residual noise. The difference between eqs.\,\ref{B0} and \ref{B1} gives
\begin{equation}
\label{noise1}
2 \left (\vec B_n + B_c\hat z \right) \cdot \left ({\bf J} \Delta \vec x \right)= \Delta(B_1^2).
\end{equation} 
Solving eq.\, \ref{B0} for $B_c$  in the hypothesis $\vec B_0 = B_0 \hat z$ gives 
\begin{equation}
\label{Bc}
B_c = -B_0  -\vec B_n \cdot \hat z + B_0\left( 1- \frac{B_{n\perp}^2}{B_0^2}   \right)^{1/2}\approx -\vec B_n \cdot \hat z - \frac{B_{n\perp}^2}{2B_0}
\end{equation}
where $ B_{n\perp}$ is the projection of the noise field on the $xy$ plane. Thus, for small noise, in the first order approximation, the first parenthesis in eq.\,\ref{noise1} reduces to $2 \vec B_\perp \cdot {\bf J} \Delta \vec x$. In conclusion, using a scalar sensor to compensate homogeneous field noise in a dual channel setup  requires  accurate preliminary compensation of the gradient of the field components perpendicular to the bias field, in our case $\partial B_x / \partial x$ and $\partial B_y / \partial x$. The inhomogeneity of the parallel component $\partial B_z / \partial x$ has a direct (i.e. first-order) effect on the field modulus (and thus must be compensated to force the secondary arm at the center of its resonance) but has no first-order effects on the noise compensation efficiency.

\section{Performance in application}
\label{example}
\begin{figure}
  \includegraphics[width=8cm]{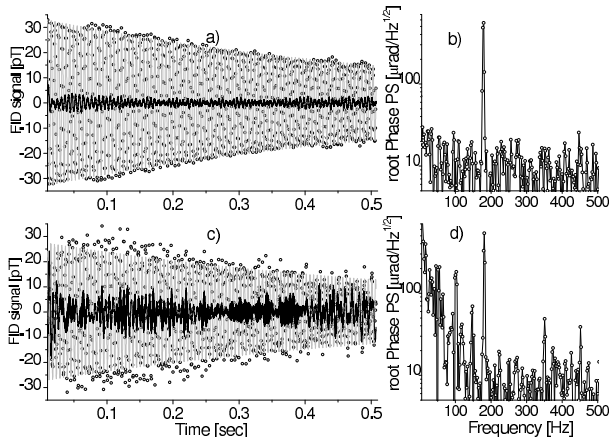}
  \caption{\label{fidsignals} The plots on the left show FID signals (dots), fitting curves (thin grey lines), and residuals (thick lines), in the time domain. The corresponding spectra are reported on the right. The upper plots are obtained with active field stabilization, while the lower plots are obtained by tracking the field and applying the methods described in \cite{noijmr09}. Time-domain traces were improved (as in \cite{noijmr09}) by identification and subtraction of components at 50\,Hz and its multiples, and linear filtering ($1^{st}$ order  Bessel filter,  170$\div$190\,Hz bandpass).
 }
\end{figure}

The compensation system was used in an ultra-low-field NMR experiment, with measurements similar to those reported in \cite{noijmr09}. Nuclear precession of remotely polarized water protons was detected in a $\mu$T field. Fig.\,\ref{fidsignals} shows the free induction decay (FID) signals and their power spectral density, obtained after averaging  700 traces. 

Active compensation of the bias field renders unnecessary the frequency rescaling procedure 
described in \cite{noijmr09},  where NMR traces were interpolated and time-rescaled to compensate the nuclear Larmor frequency drift, which was inferred from  the trace-by-trace measured bias magnetic field.

Furthermore,  operating in locked-field conditions improves the quality of the average trace, as can  easily be recognized by comparing the residual traces obtained as a difference of the average signals with their best fits (modeled as an exponentially decaying sinusoid). 
In the best-fit examples shown in Fig.\,\ref{fidsignals}, activating the field stabilization system leads to a reduction of the minimum $\chi^2$ by a factor of 7. In addition, locking the field significantly reduces  the low frequency noise, as can be clearly seen in the plot (d).

\section{Conclusion}

We have presented and characterized a simple, economic and reliable magnetic field compensation system. The system is integrated in a set-up working at a bias field of a few $\mu$T, which is measured by means of a dual-channel magnetometer, based on all-optical atomic scalar sensors and operating as a self-oscillator. 

The compensation system is driven by one of the two channels  and acts on the field component parallel to the bias, feeding back the current supplied to a Helmholtz pair. It counteracts the common mode magnetic field fluctuations, thus facilitating  difference-mode detection. The difference-mode signal reproduces the field anomalies generated by a weak source located in the proximity of one sensor.

The operating principle makes the compensation partial, meaning that only the
modulus of the magnetic field is kept constant. The limitations posed by this feature have been discussed, arriving at the conclusion that 
the compensation efficiency is mainly degraded by inhomogeneities of the field components perpendicular to the bias direction, which therefore require careful control.

Detailed data analyses were carried out to characterize the short-term performance of the stabilization system and to evaluate the phase noise spectra, from which the stray field extinction ratio was inferred. Concerning the power-net disturbances, we demonstrated an extinction factor in excess of 40\,dB. 

As a practical example, we reported results obtained with and without active compensation in an ultra-low-field NMR experiment. The free induction decay of remotely polarized water samples was recorded and averaged over long  intervals. The possibility of using these NMR measurements  to assess the long term performance of the compensation system was also considered.

\begin{acknowledgments}
The authors are grateful to Emma Thorley of the UniSi Translation Service for reviewing the manuscript.
\end{acknowledgments}

\end{document}